\documentclass[useAMS,usenatbib]{mn2e}

\usepackage{epsf}

\title[A new type of long gamma--ray burst]{A new type of long gamma--ray burst
} \author[Andrew King, Emma Olsson and Melvyn B.
Davies]{Andrew King$^1$\thanks{email:ark@astro.le.ac.uk}, Emma
Olsson$^1$ and Melvyn B. Davies$^2$ \\$^{1}$ Theoretical Astrophysics
Group, University of Leicester, Leicester LE1 7RH, UK.\\ $^{2}$ Lund
Observatory, Box 43, SE--221 00, Lund, Sweden.}

\begin{document}

\date{Accepted;Received;}

\pagerange{\pageref{firstpage}--\pageref{lastpage}} \pubyear{2006}

\maketitle
\def\msun{{\rm M_{\odot}}}

\label{firstpage}
\begin{abstract}
We consider gamma--ray bursts produced by the merger of a massive
white dwarf with a neutron star. We show that these are likely
to produce long--duration GRBs, in some cases definitely without an
accompanying supernova, as observed recently. This class of burst
would have a strong correlation with star formation, and occur close
to the host galaxy. However rare members of the class need not be near
star--forming regions, and could have any type of host galaxy. Thus a
long--duration burst far from any star--forming region would also be a
signature of this class. Estimates based on the existence of a known
progenitor suggest that our proposed class may be an important
contributor to the observed GRB rate.
\end{abstract}


\section{Introduction}

There is now good reason to believe that a large fraction of
gamma--ray bursts (GRBs) occur when rapidly--rotating massive stars
end their lives and collapse to form black holes \citep{ref1}. In some
cases this event is accompanied by a supernova explosion
\citep{ref2,ref3,ref4}. The dynamical timescale of the collapsing core
sets a lower limit to the GRB timescale, and suggests that such bursts
must have long duration ($\ga 2$~s). Much of the recent rapid
observational progress in this field comes from studying the
afterglows which are thought to occur when matter expelled in the
burst collides with its environment, which may well be matter lost by
the star at earlier times. This `collapsar' picture naturally predicts
a close association between long--duration GRBs and star formation, as
is observed. In a collapsar the accretion energy release is probably
accompanied by explosive nuclear burning, which may expel the outer
layers of the star as a supernova. While these would be undetectable
at high redshift, supernovae are indeed detected in most nearby
long--duration bursts.

GRBs with short ($\la 2$~s) durations must instead involve the
collapse or disruption of a much more compact object such as a neutron
star. A plausible picture is the merger of a pair of neutron stars, or
of a neutron star and a black hole \citep{ref5,ref6,ref7,ref8,ref9},
brought about by gravitational wave emission. Because the merger may
be long delayed after their formation, such systems are neither
associated with supernovae nor with star formation. Indeed the high
space velocities they may acquire at the formation of one or both
compact objects mean that many have travelled significant distances
from their host galaxies before merging. This in turn means that there
is little matter around the merger, reducing the afterglow brightness
and suggesting a generally harder radiation spectrum. This may also be
the reason why short--duration bursts show no energy--dependent lag in
their emission, whereas almost all long--duration bursts show a
correlation between peak energy and time.  This is usually interpreted
in terms of a more relativistic outflow, corresponding to less matter
along to rotation axis in a binary merger. A supernova explosion does
not occur in the merger picture of short GRBs, because the nuclear
energy release per unit mass (only $6\times 10^{18}$~erg\, g$^{-1}$
even for hydrogen burning, and considerably less for heavier nuclei)
is far smaller than accretion yield ($\ga 10^{20}$~erg\, g$^{-1}$),
which is also the amount needed to expel matter to infinity from the
vicinity of the accreting neutron star or black hole.

The current understanding of GRBs based on these two models thus
arrives at a fairly straightforward dichotomy, in which long bursts
are associated with star formation and sometimes with supernovae, and
have bright afterglows, softer spectra and energy--dependent lags. By
contrast short bursts should have no relation to recent star
formation, be generally displaced from their hosts, and have no
associated supernovae, fainter afterglows, harder spectra, and no
energy--dependent lags.

While this picture has generally held up well, there are recent signs
that it may not be the entire story. In particular GRB 060614
\citep{ref10,ref11,ref12} is long ($\sim 100$~s), relatively nearby
(redshift $z = 0.125$), but has no evidence of any accompanying
supernova, which would have to be more than 100 times fainter than any
ever observed. It also shows no energy--dependent time lag in its
emission.

Here we suggest that such long GRBs without supernovae or lags may be
one possible result of the merger of a neutron star and a massive
white dwarf. This class of GRB is the natural outcome of a known
evolutionary channel, which should contribute a significant fraction
of long--duration bursts. The majority of these NS+WD mergers must be
associated with star formation and lie close to the host, but a subset
can occur in any type of host galaxy. Our proposed GRB channel may be
very common, as the absence of a supernova can only be established in
nearby bursts.

\section{Merger models for GRBs}

Were it not for its long duration ($\sim 100$~s), GRB 060614 would
closely conform to the main expectations of the usual merger picture
of short GRBs sketched above. We therefore re--examine such merger
models.

The basic ingredient of mergers is unstable mass transfer, i.e. the
tidal lobe of the mass donor shrinks relative to that star's radius,
as a direct consequence of mass transfer itself. This occurs typically
when the donor/accretor mass ratio exceeds a critical value close to
unity, and is a runaway process, which develops over a few times the
orbital period $P$ of the binary. A donor star fills its tidal lobe if
and only if the orbital period is comparable with its dynamical time
$t_{\rm dyn}\sim (R^3/GM)^{1/2}$, where $M, R$ are its mass and
radius, so this is often called the dynamical instability, and it
produces typical peak mass transfer rates $\dot M \sim M/P$. Much of
the star is smashed up into a torus surrounding the accretor. From
angular momentum conservation the torus has a similar scale to the
pre--instability orbit, which is itself a few times the radius of the
donor. The rate at which mass now lands on the NS or BH accretor is
set by the torus. As this is self--gravitating and thus subject to
nonaxisymmetric gravitational instabilities, the dynamical timescale
is a reasonable estimate for the burst duration, as in a collapsar.

However, dynamical instability is not confined to neutron star
donors. A white dwarf donor is also subject to the instability (see
e.g. \citet{ref23}, \citet{ref13}) if it has mass
$M_{\rm wd} \ga 0.66M_{\rm accretor}$. Since $M_{\rm wd}$ cannot
exceed the Chandrasekhar mass $\simeq 1.4\msun$ this requires $M_{\rm
accretor} \la 2.1\msun$, and suggests that the process is most likely
with a neutron--star accretor. (Note however that low--mass black
holes may also be possible. \citet{ref25} suggest
that the high--mass X--ray binary 4U 1700--37 may contain a black hole
of $\sim 2\msun$.) For a neutron star accretor mass $1.4\msun$ we
require $M_{\rm wd} \ga 0.9\msun$. A white dwarf has $R \simeq
10^9$~cm, and fills its tidal lobe in a binary with period $P \sim
10$~s, the precise value depending on its mass. As before, the white
dwarf is probably totally disrupted into a torus, this time with
lengthscale $\sim 10^9$~cm. As this is if anything larger than the
torus likely in the collapsar picture, one would expect a similar or
longer dynamical time, and thus a long GRB. As argued previously for
NS+NS mergers, supernovae are unlikely if most nuclear burning occurs
near the neutron star, as the nuclear energy yield is far too low to
expel matter, although we shall consider a possible exception to this
later. We conclude that the unstable merger of a massive white dwarf
and a neutron star can produce long GRBs without accompanying
supernovae.

Several NS + massive WD binaries are known in the Galaxy, as the
neutron stars can be detected as pulsars. The most promising object is
PSR J1141--6545 \citep{ref14}, which has a 5~hr orbit and will merge
in about $10^9$~yr under the effect of gravitational
radiation. Because the orbit is relativistic, the masses are known to
great precision \citep{ref16} as $M_{\rm wd} = 0.986 \pm 0.02 \msun$
and $M_{\rm ns} = 1.30 \pm 0.02 \msun$. This firmly establishes the
mass ratio as $M_{\rm wd}/M_{\rm ns} > 0.73$, making dynamical
instability and a GRB inevitable.

\citet{ref15} discuss the formation process for PSR J1141--6545 in
detail. Starting from a pair of main sequence stars, the more massive
primary fills its Roche lobe as a giant and transfers its mass to the
companion, which then becomes more massive than the primary was
originally. The core of the primary becomes a massive white dwarf,
while the companion is now massive enough to become a helium
star. This star eventually fills its Roche lobe and transfers its
envelope to the white dwarf at a high rate, causing most of it to be
ejected from the binary, which shrinks. Ultimately the helium star
explodes as a supernova, producing a neutron star in the required
tight orbit with the massive white dwarf. \citet{ref15} estimate a
formation rate of $5\times 10^{-5} - 5 \times 10^{-4}$~yr$^{-1}$ for
such systems in the Galaxy (their Fig. 8) and further show that over
half of them merge within $10^8$~yr, and 95\% within a Hubble time
(their Fig. 10).

This suggests that such systems can indeed make a significant
contribution to the gamma--ray burst rate in the
Universe. \citet{ref15} also compute the distribution of supernova
recoil kick velocities for these systems (their Fig. 9). A substantial
fraction have relatively small velocities $\la 100$~km~s$^{-1}$. This
is promising for models of GRB 060614, as this and the merger
timescale distribution suggest that in many cases the GRB must occur
with the system still relatively close to the host galaxy, as observed
for GRB 060614 even though the host is a dwarf. Together with the high
initial merger rate (100 times higher in the first $10^8$~yr) this
shows that NS + massive WD bursts must show a strong correlation with
star formation, although the relatively rare later mergers would show
no such correlation.

We have mentioned that an accompanying supernova is unlikely if
nuclear burning occurs close to the neutron star. The one possible
escape from this conclusion would occur if it were possible to
initiate burning at much larger distances. We note that since the
merger contains no H or He such supernovae would automatically be of
Type Ic, as indeed observed for GRB supernovae. However this possible
supernova mechanism is probably ruled out for white dwarfs
sufficiently massive to have ONeMg rather than CO composition ($M_{\rm
wd} \ga 1.1\msun$), because neutrino energy losses through electron
capture (particularly on Mg) (see e.g \citet{ref18}) prevent explosive
burning.

We note that \citet{ref17} suggested that NS + WD mergers might not
produce GRBs. They reasoned that the neutron star might build up a
spherically symmetric atmosphere, which could make a spherical
explosion from the surface baryon--rich, and therefore
sub--relativistic. This is evidently a less serious problem for an
explosion directed along the rotational axis, as is now usually
assumed to be the case. A related question is whether the accretor
survives as a neutron star rather than collapsing to a black hole
early in the hyperaccretion phase. The total mass $\simeq 2.3\msun$ of
PSR J1141--6545 is lower than some theoretical estimates of the
maximum neutron--star mass (e.g. \citet{ref20}) and probably
less than the mass potentially reached by neutron stars in some
low--mass X--ray binaries, but calculations of NS+NS mergers generally
lead to early collapse to a black hole (e.g. \citet{ref22}). The accretion yield on a neutron star is $\sim 10$\%,
comparable with all but the most rapidly--spinning black holes
(42\%). Evidently there is sufficient energy release to power a
gamma--ray burst in both cases, but a late collapse to a black hole
would presumably cause a second energetic event (similar to the
supranova model of \citet{ref24}, where angular momentum loss
rather than continued accretion drives the collapse). This is
conceivably interesting in connection with the late flares seen in
some GRBs, e.g.  GRB 050502b \citep{ref19}). One possible
discriminant between the two cases might be the properties of the jets
needed to make the gamma--rays, but we note that neutron--star X--ray
binaries are apparently able to make jets too (cf \citet{ref21}).

\section{Conclusions}

We have suggested that GRB 060614 is representative of a new class of
gamma--ray burst in which a massive white dwarf merges with a neutron
star. The characteristics of this class are clear. They are
long--duration GRBs, in some cases definitely without an accompanying
supernova, which may show some other features usually associated with
short bursts, such as a lack of energy--dependent time lags and
perhaps much weaker afterglows.

As a class these bursts show a strong correlation with star formation,
and occur close to the host galaxy. However rare members of the class
need not so correlate, and can have any type of host galaxy. Thus a
detection of a long--duration burst far from any star--forming might
be a signature of one of these bursts. 

Our estimates based on PSR J1141-6545 suggest that this proposed type
of GRB may provide an important fraction of the observed GRB rate. We
note that the absence of a supernova can only be established in nearby
bursts.

\section*{Acknowledgments} ARK gratefully acknowledges a Royal
Society--Wolfson Research Merit Award. EO is supported by a European
Union Research and Training Network grant. MBD is a Royal Swedish
Academy Research Fellow supported by a grant from the Knut and Alice
Wallenberg Foundation.

\label{lastpage}

\end{document}